
\magnification=\magstep1
\hsize 6.0 true in
\vsize 9.0 true in
\voffset=-.5truein
\pretolerance=10000
\baselineskip=22truept

\font\tentworm=cmr10 scaled \magstep2
\font\tentwobf=cmbx10 scaled \magstep2

\font\tenonerm=cmr10 scaled \magstep1
\font\tenonebf=cmbx10 scaled \magstep1

\font\eightrm=cmr8
\font\eightit=cmti8
\font\eightbf=cmbx8
\font\eightsl=cmsl8
\font\sevensy=cmsy7
\font\sevenm=cmmi7

\font\twelverm=cmr12
\font\twelvebf=cmbx12
\def\subsection #1\par{\noindent {\bf #1} \noindent \rm}

\def\mid {\let\rm=\tenonerm \let\bf=\tenonebf \rm \bf}

\def\para{\par \vskip 12 pt}

\def\head{\let\rm=\tentworm \let\bf=\tentwobf \rm \bf}

\def\heading #1 #2\par{\centerline {\head #1} \smallskip
 \centerline {\head #2} \vskip .15 pt \rm}

\def\eight{\let\rm=\eightrm \let\it=\eightit \let\bf=\eightbf
\let\sl=\eightsl \let\sy=\sevensy \let\m=\sevenm \rm}

\def\foots{\noindent \eight \baselineskip=10 true pt \noindent \rm}
\def\sexion{\let\rm=\twelverm \let\bf=\twelvebf \rm \bf}

\def\section #1 #2\par{\vskip 20 pt \noindent {\mid #1} \enspace {\mid #2}
  \para \noindent \rm}

\def\abstract#1\par{\para \foots {\bf Abstract: \enspace}#1 \para}

\def\author#1\par{\centerline {#1} \vskip 0.1 true in \rm}

\def\abstract#1\par{\noindent {\bf Abstract: }#1 \vskip 0.5 true in \rm}

\def\sqr#1#2{{\vcenter{\vbox{\hrule height.#2pt
  \hbox {\vrule width.#2pt height#1pt \kern#1pt
  \vrule width.#2pt}
  \hrule height.#2pt}}}}

\def\n{\noindent}
\def\s{\smallskip}
\def\m{\medskip}
\def\b{\bigskip}
\def\c{\centerline}

\def\gne #1 #2{\ \vphantom{S}^{\raise-0.5pt\hbox{$\scriptstyle #1$}}_
{\raise0.5pt \hbox{$\scriptstyle #2$}}}

\def\ooo #1 #2{\vphantom{S}^{\raise-0.5pt\hbox{$\scriptstyle #1$}}_
{\raise0.5pt \hbox{$\scriptstyle #2$}}}


\voffset=-.5truein
\vsize=9truein
\baselineskip=22pt
\hsize=6.0truein
\pageno=1
\pretolerance=10000
\def\n{\noindent}
\def\s{\smallskip}
\def\b{\bigskip}
\def\m{\medskip}
\def\c{\centerline}

\line{\hfill IUCAA - 18/95}

\line{\hfill June 1995}
\m
\m

\c{\bf\mid A CLASS OF STATIONARY ROTATING }
\c{\bf\mid STRING COSMOLOGICAL MODELS}
\b
\b
\b
\n L.K. Patel$^{1,3} $ \& Naresh Dadhich$^{2,*} $
\b
\item{1.} Department of Applied Mathematics, University of Zululand,
Private bag X1001, kwa-Dlangezwa 3886 South Africa.
\b
\item{2.} Inter-University Centre for Astronomy and Astrophysics, Post Bag 4,
Ganeshkhind, Pune - 411 007.
\b
\b
\c{\bf Abstract}

\vskip 0.35 cm

We obtain a one parameter class of stationary rotating string cosmological
 models of which the well-known G$\ddot o $del Universe is a particular case.
By suitably choosing the free parameter function, it is always possible to
satisfy the energy conditions. The rotation of the model hinges on the
cosmological constant which turns out to be negative.

\b
\b
\s
\n {\bf PACS numbers :} 04.20 Jb, 98.80 Dr
\s
\n {\it Key words :} General relativity, exact solutions, rotating string
cosmology.
\b
\b
\b
\b
\b
\s
\baselineskip=9pt
\n $^3 $ Permanent address : Department of Mathematics, Gujarat University,
\s
\n~~ Ahmedabad 380 009, India
\b
\n $^* $ E-mail: naresh@iucaa.ernet.in
\vfill\eject
\baselineskip=22pt

Cosmic strings have been considered in the study of early Universe cosmology.
 They may be one of the sources of density perturbations that are required for
formation of large scale structure in the Universe [1,2]. They possess
stress-energy and hence couple to gravitational field. Their various features
 have been considered by some authors [3-5]. Cosmic strings as source of
gravitational field in general relativity (GR) was discussed by Letelier [6]
and Stachel [7]. Letelier [8] has further constructed string cosmological
 models for Bianchi I and Kantowski-Sachs spacetimes by introducing the
energy-momentum tensor.

$$ T_{ik} = \rho u_i u_k - \lambda w_i w_k,~ u_i u^i = 1 = -w_i w^i,~ u_i w^i =
0 \eqno (1) $$

\n as the source term in Einstein's equation

$$ R_{ik} - {1 \over 2} R g_{ik} = -8 \pi T_{ik} - \wedge g_{ik} \eqno (2) $$

\n where $\wedge $ is the cosmological constant. $T_{ik} $ represents the
 energy momentum of a cloud of strings attached with mass particles. The
 density $\rho $ is made up of particle density $\rho_p $ and the string
tensor $\lambda $, and is given by

$$ \rho = \rho_p + \lambda \eqno (3) $$

\n The energy conditions will require $\rho \geq 0 $,~ $\rho_p \geq 0 $
leaving the sign of $\lambda $ undetermined. $\lambda $ has however to be
 positive whenever $\rho_p = 0 $. The matter flow and the string fibre
directions are respectively specified by the unit timelike $u^i $ and
 spacelike $w^i $ vectors.
\s
String cosmological models have been studied for Bianchi type spacetimes
by several authors [9-13]. It is also shown that cylindrically symmetric
non-singular spacetimes also admit physically reasonable string
cosmological models [14]. So far all the models considered are free of
rotation and non-static. In this paper we obtain stationary rotating
string solutions of Einstein's equation. We have obtained a one parameter
class of rotating string spacetimes and the free function can be suitably
 chosen to satisfy the energy conditions. The well-known rotating G$\ddot o$del
Universe follows as a special case of this class. It turns out that
the cosmological constant $\wedge $ plays a very simportant role in the
sense that it measures rotation as well as particle density $\rho_p $.
\s
We consider the stationary line-element in the form

$$ ds^2 = -dx^2 - \alpha^2 dy^2 - dz^2 + (dt + H dy)^2 \eqno (4) $$

\n where $\alpha $ and $H $ are functions of $x $ alone. We introduce
 the orthonormal tetrad; $\theta^1 = dx,~\theta^2 = \alpha dy ~ \theta^3
= dz $ and $\theta^4 = dt + H dy $ and in what follows all quantities will
 be referred to the tetrad frame.
\s
The surviving $R_{ab} $ are given as follows :

$$  R_{11} = R_{22} = \alpha^{\prime \prime}/\alpha  -  H^{\prime
2}/2 \alpha^2 $$

$$  R_{44}  = -H^{\prime 2}/2 \alpha^2,~ R_{24} =  -(1/2  \alpha)
(H^{\prime \prime} - H^{\prime} \alpha^{\prime} / \alpha ). \eqno
(5) $$

\n Substituting this in (2) and using (1), we get

$$ R_{24} = 0 \eqno (6) $$

$$ R_{11} = -4 \pi (\rho + \lambda) - \wedge \eqno (7) $$

$$ R_{44} = 2 \wedge = 8 \pi (\lambda - \rho ) \eqno (8) $$

\n  where we have used $u_i = \delta_i^4 $ and $w_i =  \delta^3_i  $
(string is along the z-axis).
\s
{}From eqns.(5)-(8) we readily obtain

$$  H^{\prime}  = m \alpha,~ 8 \pi \rho =  m^2  -  \alpha^{\prime
\prime}/\alpha,~   8  \pi  \lambda  =  m^2/2   -   \alpha^{\prime
\prime}/\alpha \eqno (9) $$

\n where $m $ is a constant of integration. Clearly $\rho \geq 0 $
is  ensured if $\alpha^{\prime \prime} \leq 0 $ and the  particle
density

$$ \rho_p = \rho - \lambda = m^2/2 \geq 0 \eqno (10) $$

\n whereas the cosmological constant

$$ \wedge = -m^2/4 \leq 0. \eqno (11) $$

\n  The  vorticity of the velocity field, $\Omega =  w_{ab}  w^{ab}  $
turns out to be

$$  \Omega   = \sqrt 2 m \eqno (12) $$

\n  which  will vanish only when $\wedge = 0 $ and  so  does  the
particle density $\rho_p $.
\s
The  metric function $\alpha $ is undetermined and hence we  have
one  parameter  class  of  rotating  string  spacetimes.  In  the
followng we consider some simple interesting cases.
\s
\n {\bf Solution 1:} The simplest case will obviously be  $\alpha
= x $ leading to

$$  \lambda = \rho_p = 2 \rho = m^2/16 \pi,~ H = m  x^2/2,  \eqno
(13) $$

\n Then the metric reads

$$  ds^2 = -dx^2 -x^2 dy^2 -dz^2 + (dt + {1 \over 2} mx^2  dy)^2.
\eqno (14) $$
\s
\n{\bf  Solution  2:} Let us put $\lambda=0 $  which  will  imply
$\alpha = e^{mx/\sqrt 2},~ H = \sqrt 2 \alpha $. Then we obtain

$$  ds^2  = -dx^2 - e^{\sqrt 2 mx} dy^2 - dz^2 + (dt  +  \sqrt  2
e^{mx/\sqrt 2} dy)^2 \eqno (15) $$

\n  which is the well-known G$\ddot o $del Universe [15] with  $8
\pi \rho = m^2/2 = -2 \wedge $.
\s
\n {\bf Solution 3:} Let us consider the equation of state of the
kind $\rho = (1 + k) \lambda
                            $ where $k $ is a positive constant. Then  we
have

$$ \alpha^{\prime \prime} + {1-k \over 2k} m^2 \alpha = 0 $$

\n the solution of which depends upon the sign of ${1 - k \over 2k} $.

\n Case (i) $a^2 = {1 - k \over 2k} m^2 > 0 $. In that case,

$$ \alpha = cos a x,~ 8 \pi \rho = {m^2 (1 + 3k) \over 2k},~ H  =
-{m \over a} sin a x $$

\n and the metric reads

$$ ds^2 = -dx^2 -cos^2 ax dy^2 - dz^2 + (dt - {m \over a} sin  ax
dy)^2. \eqno (16) $$

\n Case (ii) $-b^2 = {1-k \over 2k} m^2 $. We then obtain

$$ \alpha = e^{bx},~8 \pi \rho = {1 + k \over 2k},~ H = {m  \over
b} e^{bx} $$

\n and

$$  ds^2 = -dx^2 -e^{2bx} dy^2 - dz^2 + (dt + {m \over b}  e^{bx}
dy)^2. \eqno (17) $$

\s
\n{\bf Solution 4:} Let us consider the case of vanishing $\wedge
$  which means $\rho_p = \Omega =  0 $ and $\rho = \lambda $.
This  is the  case of the Universe filled with the cosmic  strings
 alone. Note  that $\alpha $ still remains free to be chosen.  We  choose
$\alpha^{\prime  \prime}/\alpha = -n^2 $, $n $ being  a  constant.
Then we get

$$ \rho = \lambda = n^2/8 \pi $$

\n and the metric has the simple form

$$ ds^2 = -dx^2 - cos^2 nx dy^2 - dz^2 + dt^2. \eqno (18) $$

\n  It  is  interesting to note that $\rho $ is  a  constant  and
switching that off leads to flat spacetime.
\s
This metric was earlier obtained by Patel and Vaidya [16] and was
interpreted as magnetic Universe with cosmological constant. That
is   the   same  spacetime  can  have  two   different   physical
visualisations.
\s
Finally  we would like to mention that all the  cases  considered
above  satisfy  the energy conditions and  hence  are  physically
admissible.  Further  they can always be  satisfied  by  suitably
choosing the free function $\alpha $. It would be interesting  to
find non-static rotating string models.
\s
\n {\bf Acknowledgement :} LKP wishes to thank the University  of
Zululand  for  hospitality  and the  University  of  Gujarat  for
granting the leave of absence.

\vfill\eject
\n References :
\item{[1]} T W S Kibble (1976) J. Phys. {\bf A 9}, 1387.
\s
\item{[2]} Ya B Zeldovich (1980) Mon. Not. R. Astron. Soc. {\bf 192}, 663.
\s
\item{[3]} A Vilenkin (1981) Phys. Rev. {\bf D 24}, 2082.
\s
\item{[4]} J R Gott (1985) Astrophys. J. {\bf 288}, 422.
\s
\item{[5]} D Garfinkle (1985) Phys. Rev. {\bf D32}, 1323.
\s
\item{[6]} P S Letelier (1979) Phys. Rev. {\bf D20}, 1294.
\s
\item{[7]} J Stachel (1980) Phys. Rev. {\bf D 21}, 2171.
\s
\item{[8]} P S Letelier (1983) Phys. Rev. {\bf D 28}, 2414.
\s
\item{[9]} K D Krori, T Chaudhuri, C R Mahanta and A Mazumdar (1990)
Gen. Rel. Grav. {\bf 22}, 123.
\s
\item{[10]} A Banerjee, A K Sanyal and S Chakraborty (1990) Pramana -
J.Phys. {\bf 34}, 1.
\s
\item{[11]} R Tikekar and L K Patel (1992) Gen. Rel. Grav. {\bf 24}, 397.
\s
\item{[12]} R Tikekar and L K Patel (1994) Pramana - J.Phys. {\bf 42}, 483.
\s
\item{[13]} S D Maharaj, P G L Leach and K S Govinder (1995) - to appear in
Pramana.
\s
\item{[14]} R Tikekar, L K Patel and N Dadhich (1994) Gen. Rel. Grav. {\bf 26},
647.
\s
\item{[15]} K G$\ddot o$del (1949) Rev. Mod. Phys. {\bf 21}, 447.
\s
\item{[16]} L K Patel and P C Vaidya (1971) Current Science {\bf 40}, 278.

\bye